# Random Stable Matchings

**Stephan Mertens** †

Institut für Theoretische Physik, Otto-von-Guericke Universität, PF 4120, 39016 Magdeburg, Germany

**Abstract.** The stable matching problem is a prototype model in economics and social sciences where agents act selfishly to optimize their own satisfaction, subject to mutually conflicting constraints. A stable matching is a pairing of adjacent vertices in a graph such that no unpaired vertices prefer each other to their partners under the matching. The problem of finding stable matchings is known as stable marriage problem (on bipartite graphs) or as stable roommates problem (on the complete graph). It is well-known that not all instances on non-bipartite graphs admit a stable matching. Here we present numerical results for the probability that a graph with $n$ vertices and random preference relations admits a stable matching. In particular we find that this probability decays algebraically on graphs with connectivity $\Theta(n)$ and exponentially on regular grids. On finite connectivity Erdös-Rényi graphs the probability converges to a value larger than zero. Based on the numerical results and some heuristic reasoning we formulate five conjectures on the asymptotic properties of random stable matchings.

## 1. Introduction

The stable matching problem is a prototype model in economics and social sciences where agents act selfishly to optimize their own satisfaction, subject to mutually conflicting constraints. The best known example is the stable marriage problem [1, 2], where the agents are $n$ men and $n$ women that compete with each other in the "marriage market". Each man ranks all the women according to his individual preferences, and each woman does the same with all men. Everybody wants to get married to someone at the top of his or her list, but mutual attraction is not symmetric and frustration and compromises are unavoidable. A minimum requirement is a matching of men and women such that no man and woman would agree to leave their assigned partners in order to marry each other. Such a matching is called stable since no individual has an icentive to break it.

The stable marriage problem was introduced by David Gale and Lloyd Shapley in 1962 [3]. In their seminal paper they proved that each instance of the marriage problem has at least one stable solution, and they presented an efficient algorithm to find it. The Gale-Shapley algorithm has been applied to many real-world problems, not by dating agencies but by central bodies that organize other two-sided markets like the assignment of students to colleges or residents to hospitals [4]. Besides its practical relevance, the stable marriage problem has many interesting theoretical features that have attracted researchers from computer science, mathematics, economics, game theory, operations research and–more recently–physics [5, 6, 7, 8, 9].

† stephan.mertens@physik.uni-magdeburg.de.

The salient feature of the stable marriage problem is its bipartite structure: the agents form two groups (men and women), and matchings are only allowed between these groups but not within a group. This is adequate to describe two-sided markets like the assignment of students to colleges or residents to hospitals [4], but in other applications there is only one group of agents that want to matched to each other. Examples are the formation of cockpit crews from a pool of pilots or the assignment of students of the same sex to the double bedrooms in a dormitory. The latter is known as the stable roommates problem and was also introduced by Gale and Shapley [3]. Gale and Shapley presented a small example to demonstrate an intriguing difference between the marriage and the roommates problem: Whereas the former always has a solution, the latter may have none. Here is the example:

$$\begin{array}{cccc} 1: & 3 & 2 & 4 \\ 2: & 1 & 3 & 4 \\ 3: & 2 & 1 & 4 \\ 4: & 1 & 2 & 3 \end{array} \tag{1}$$

This table represents the preferences of 4 people. Person 1 likes person 3 in the first place, person 2 in the second place and so on. Apparently person 4 is not very popular, but someone has to share a room with him. If we match 4 with 1 and 2 with 3 than 1 is very unhappy and he will ask 2 and 3 to share a room with him. 2 will accept this offer because he can improve, too, and together they will screw up the whole arrangement. The other two possible matchings are unstable, too.

Bipartiteness is crucial for the solvability of a matching problem. The Gale-Shapley algorithm for example does not work for non-bipartite problems like the stable roommates problem. In fact some people believed that the roommates problem was NP-complete, but more than 20 years after the Gale-Shapley paper Robert Irving presented a polynomial time algorithm for the stable roommates problem [10]. Irving's algorithm either outputs a stable solution or "No" if none exists. This was a major breakthrough, but still the problem was (and is) not fully understood, see e.g. the "Open Problems" section in [1]. One of the open issues is the probability $P_n$ that an arbitrary roommates instance of size $n$ is solvable. Numerical simulations indicate that $P_n$ is a monotonically decreasing function of $n$, but the data "... is not really conclusive enough to add support to any strong conjecture as to the ultimate behavior of $P_n$" [11]. In this contribution we will present numerical data that *is* conclusive enough to conjecture the asymptotic behavior of $P_n$.

In the stable roommates problem, everybody knows (and ranks) everybody else. In real world problems the agents do not have that much information. Their situation corresponds more to the stable matching problem in finite connectivity graphs. We will investigate $P_n$ for two types of graphs with finite connectivity: regular lattices and random graphs. In both cases numerical data and heuristic arguments are sufficient to conjecture the asymptotic behavior of $P_n$.

Before we present the results we will define the stable matching problem on general graphs and discuss a certificate for the solvability of an instance.

## 2. Stable Matchings and Stable Permutations

An instance of the stable matching problem is completely specified by a preference table $T$. $T$ has a row for each agent, and agent $v$'s row contains all the other agents that $v$ accepts as partners, linearly ordered according to agent $v$'s preferences. Throughout the paper we will assume that the order is strict (no ties) and that acceptance is mutual, i.e. $w$ is in $v$'s list if and only if $v$ is in $w$'s list. The preference table can be interpreted as the adjacency list of a graph $G = (V, E)$ in which the agents $v$ are the nodes and $\{v, w\} \in E$ if and only if $v$ and $w$ accept each other as partners. See figure 1 for an example.

**Figure 1.** Example of a stable matching problem: acceptability graph $G$ (left) and preference table $T$ (right). The matching indicated by blue edges covers all vertices but is not stable. The red edges form a stable matching.

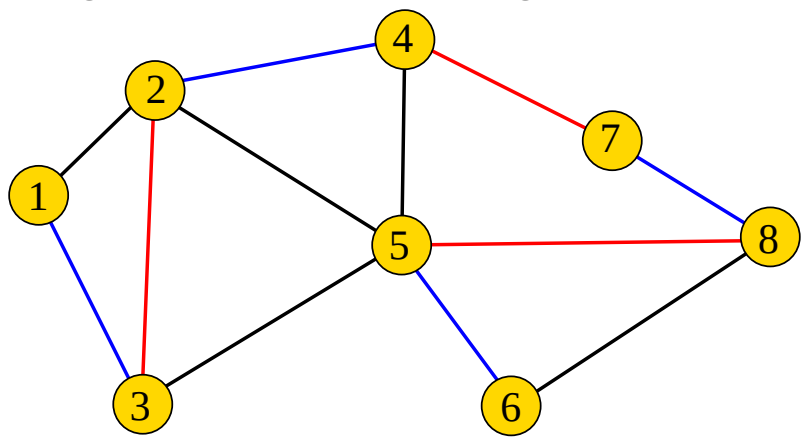

| | | | | | |
|---|---|---|---|---|---|
| 1: | 2 | 3 | | | |
| 2: | 3 | 1 | 5 | 4 | |
| 3: | 2 | 5 | 1 | | |
| 4: | 5 | 2 | 7 | | |
| 5: | 3 | 2 | 8 | 4 | 6 |
| 6: | 5 | 8 | | | |
| 7: | 8 | 4 | | | |
| 8: | 5 | 7 | 6 | | |

A matching is a subset $M \subset E$ of non-adjacent edges. A pair $v, w$ is called blocking with respect to $M$ if $\{v, w\} \in E \setminus M$ and in addition one of the following conditions holds:

(i) both $v$ and $w$ are covered by $M$ but prefer themselves to their partners in $M$, or

(ii) only $v$ is covered by $M$ and prefers $w$ to his partner in $M$ (or vice versa), or

(iii) neither $v$ nor $w$ are covered by $M$.

If $v$ and $w$ form a blocking pair, they tend to ignore the matching $M$ and form a new pair $\{v, w\}$. A matching $M$ is called stable if there are no blocking pairs. Figure 1 shows an example. Here the matching $\{(1,3),(2,4),(5,6),(7,8)\}$ (blue edges) is not stable since it is blocked by $(2,3)$. The matching $\{(2,3),(4,7),(5,8)\}$ (red edges) is stable.

In the stable marriage problem the acceptability graph $G$ is the complete bipartite graph $K_{n:n}$. The Gale-Shapley algorithm does work on general bipartite graphs and constitutes a constructive proof that a stable matching always exists if the acceptability graph is bipartite. Non-bipartite graphs, on the other hand, do not always allow a stable matching. It is easy to construct an instance that blocks all matchings [12]: A non-bipartite graph must contain at least one cycle of odd length. Let $v_1, v_2, \ldots, v_k$ be such a cycle. We construct the preference table $T$ such that $v_i$ ranks its predecessor $v_{i-1}$ first and its successor $v_{i+1}$ second (figure 2). According to their mutual highest rankings, the members of this cycle prefer to stay among themselves. This is why we call such a cycle *exclusive*. In an exclusive cycle with an odd number of members at least one member has to find a partner outside the cycle. Let us assume that, under a matching $M$, this poor chap is $v_i$. Then $v_i$ prefers $v_{i+1}$ to its current situation (whether or not $v_i$ is covered by $M$). Since $v_i$ is $v_{i+1}$'s first choice, $v_{i+1}$ prefers $v_i$ to its current situation and $(v_i, v_{i+1})$ form a blocking pair. Hence matching $M$ is not stable.

**Figure 2.** Example of an odd exclusive cycle that prevents any matching from being stable.

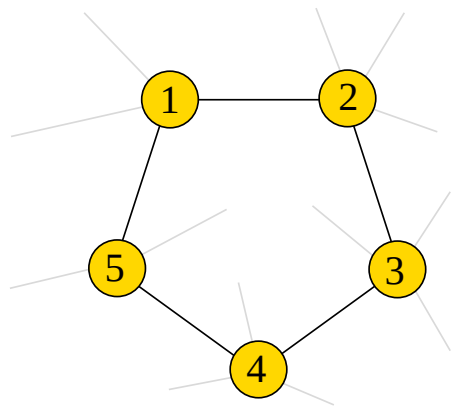

| | | | |
|---|---|---|---|
| 1: | 5 | 2 | $\cdots$ |
| 2: | 1 | 3 | $\cdots$ |
| 3: | 2 | 4 | $\cdots$ |
| 4: | 3 | 5 | $\cdots$ |
| 5: | 4 | 1 | $\cdots$ |
| $\cdots$ | $\cdots$ | $\cdots$ | $\cdots$ |

Note that this is precisely what happens in example (1), where $(1,2,3)$ is an odd exclusive cycle. Presence of an odd exclusive cycle is a sufficient but not a necessary condition for the non-existence of a stable matching, but it captures the basic idea of non-solvability.

A necessary and sufficient condition for non-solvability, a *certificate*, requires a generalization of matchings. Any matching of size $n$ can be interpreted as a permutation $\Pi$ of $\{1,\ldots,n\}$ that is completely composed of cyles of length $\leq 2$. Uncovered vertices correspond to the fixed-points of $\Pi$. An obvious generalization is to allow arbitrary permutations $\Pi$, but for that one needs to extend the definition of stability. A permutation $\Pi$ is called stable if

$$\forall i : i \text{ does not prefer } \Pi(i) \text{ to } \Pi^{-1}(i) \tag{2}$$

and

$$i \text{ prefers } j \text{ to } \Pi(i) \Rightarrow j \text{ prefers } \Pi(j) \text{ to } i \tag{3}$$

To interpret the "prefers to" relation for fixed points of $\Pi$ we simply add every agent $i$ to the very end of its own preference list. Note that for permutations with 2-cycles and fixed points only (matchings) condition (2) is trivially satisfied and condition (3) reduces to the usual "no blocking pairs" condition. Condition (2) enforces each cycle of length $\geq 3$ to have a monotonic rank ordering: every member of the cycle prefers his successor to his predecessor, and condition (3) prevents any member of the cycle to leave the cycle. In general the stability of a cycle depends on the rankings in other cycles. Exclusive cycles (Figure 2) with their mutual first and second rankings satisfy (3) automatically and independently of other rankings.

Stable permutations were introduced by Tan [13], and their significance for the stable matching problem arises from the following facts:

(i) Each instance of the stable matching problem admits at least one stable permutation.

(ii) If $\Pi$ is a stable permutation for a matching instance that contains a cycle $C = (v_1, v_2, \ldots, v_{2m})$ of even length, then replacing $C$ by the 2-cycles $(v_1, v_2), \ldots, (v_{2m-1}, v_{2m})$ or by the 2-cycles $(v_2, v_3), \ldots, (v_{2m}, v_1)$, gives another stable permutation.

(iii) If $C$ is an odd-length cycle in *one* stable permutation for a given matching instance, then $C$ is a cycle in *all* stable permutations for that instance.

These facts, proven by Tan for the stable roommates problem [13] but valid for stable matchings in general graphs, establish the cycle structure of stable permutations as certificate for the existence of a stable matching. An instance of the stable matching problem is solvable if and only if the instance admits a stable permutation with no odd cycles of length $\geq 3$.

## 3. Random Instances

For the rest of the paper we consider instances of the stable matching problem where each agent arranges their preference list independently in random order.

### *3.1. Complete Graph*

The stable matching problem on the complete graph is better known as stable roommates problem, and calculating the proportion $P_n$ of solvable random instances for the stable roommates problem is an open problem [1, problem 8]. For very small sizes this probability can be calculated exactly by exhaustive enumeration of all $[(n-1)!]^{n-1}$ instances of size $n$:

$$P_4 = \frac{26}{27} = 0.96296\ldots \qquad P_6 = \frac{181431847}{194400000} = 0.93329\ldots. \tag{4}$$

Monte Carlo simulations indicate that $P_n$ is a monotonically decreasing function of $n$. Simulations up to $n = 2000$ [11] did not settle the question as to whether $P_n$ converges to

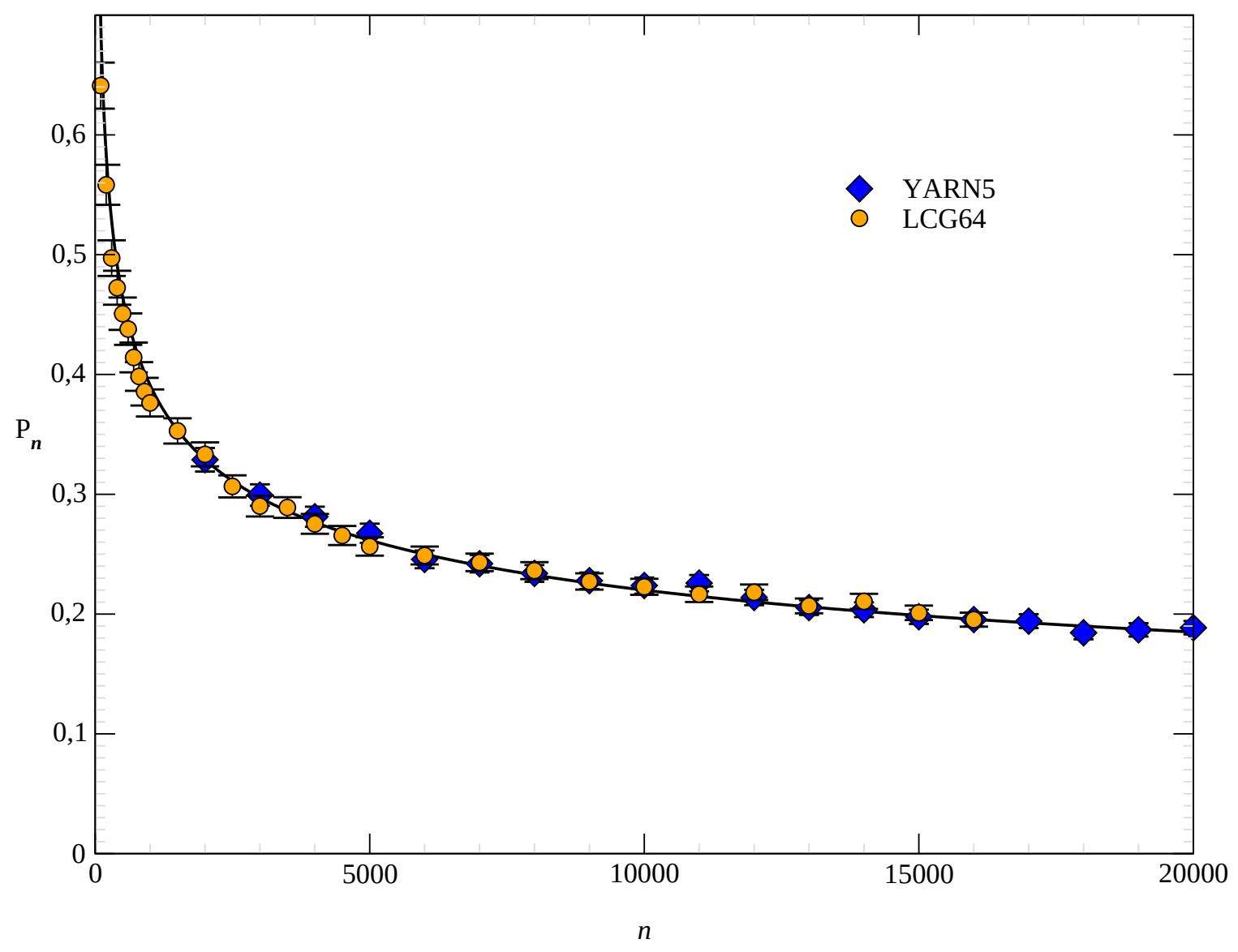

**Figure 3.** Probability $P_n$ of admitting a stable matching in the complete graph of size $n$ (stable roommates problem). Each data point represents an average over $10^4$ random samples, different symbols refer to different pseudorandom number generators from the TRNG library [14]. The line is given by equation (7).

0 or to some constant $> 0$. The problem with simulations is that the rate of convergence is rather slow. In fact Pittel [15] proved the asymptotic lower bound

$$P_n \gtrsim \frac{2e^{3/2}}{\sqrt{\pi n}} \tag{5}$$

by applying the second moment method to the number of stable matchings. An asymptotic upper bound was proven by Pittel and Irving [11],

$$\lim_{n\to\infty} P_n \le \frac{\sqrt{e}}{2} = 0.8244\ldots. \tag{6}$$

Equations (4), (5) and (6) represent all rigorously established facts on $P_n$.

We harnessed the power of a 128-CPU Linux cluster‡ to measure $P_n$ up to $n = 20000$ (figure 3). The data suggest that the true rate of convergence is even slower than (5), namely $P_n = \Theta(n^{-1/4})$. The results of our simulation are summarized in the following conjecture:

**Conjecture 1** *The probability $P_n$ that a random instance of the stable roommates problem admits a solution is asymptotically*

$$P_n \simeq e\sqrt{\frac{2}{\pi}}\, n^{-1/4}. \tag{7}$$

The conjectured algebraic decay $P_n = \Theta(n^{-1/4})$ is strongly supported by the data. The conjectured constant in (7) is a result of numerical fitting and guided guessing.

‡ http://tina.nat.uni-magdeburg.de

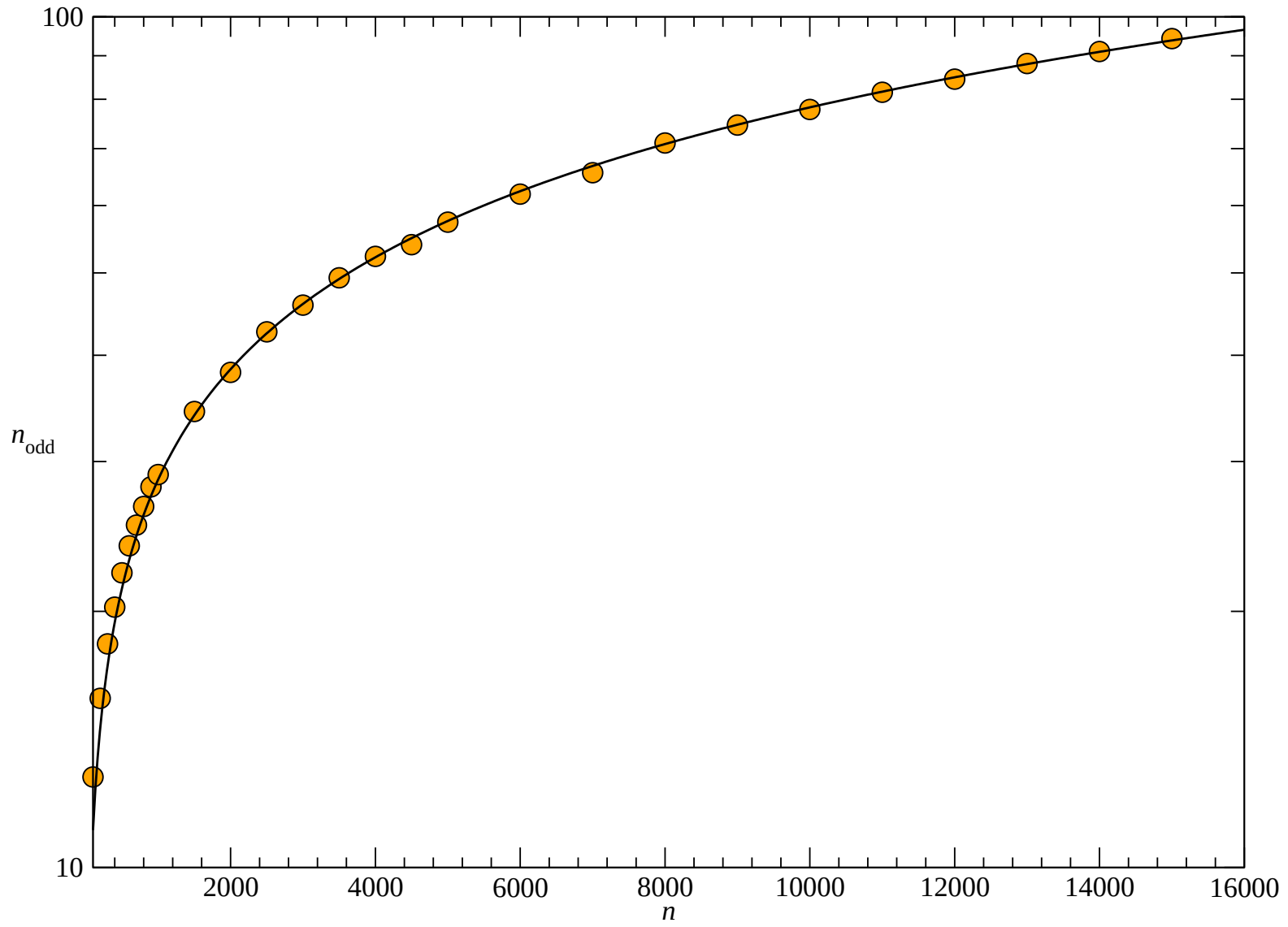

**Figure 4.** Number of agents that are elements of stable odd cycles in unsolvable stable roommates instances. Each data point represents an average over $10^4$ random samples, the line is the numerical fit $2.375\sqrt{n/\ln n}$. Note the logarithmic scaling of the ordinate.

In section 2 we have seen that every unsolvable instance of the roommates problem is characterized by a unique set of stable cycles of odd length. Let $n_{\text{odd}}$ be the total size of all odd-length cycles. In [16] it was shown that $n_{\text{odd}}$ is bounded in probability,

$$n_{\text{odd}} = \mathcal{O}_p\left(\sqrt{n \ln n}\right), \tag{8}$$

but what is the average value of $n_{\text{odd}}$? Numerical simulations (figure 4) support the following conjecture:

**Conjecture 2** *Let $n_{\text{odd}}$ be the total size of all odd-length cycles in a stable permutation of a random instance of the stable roommates instance of size $n$. Then*

$$\mathbb{E}n_{\text{odd}} = \Theta\left(\sqrt{\frac{n}{\ln n}}\right) \tag{9}$$

*where $\mathbb{E}$ denotes the average conditioned on unsolvable instances.*

The numerical constant involved in (9) is approximately 2.375.

### *3.2. Grids*

In real world applications each agent knows (and ranks) only a small subset of the other agents. A natural cause of limited information is spatial distance, i.e. each agent overlooks only their nearest neighbors. We model neighborhood by arranging the agents on point lattices with integer coordinates or *grids*. The points in an $d$-dimensional, finite grid with periodic boundary conditions are given by the set

$$\mathbb{Z}^d_{n_1 n_2 \cdots n_d} = \mathbb{Z}_{n_1} \otimes \mathbb{Z}_{n_2} \otimes \cdots \otimes \mathbb{Z}_{n_d}\,, \tag{10}$$

where $\mathbb{Z}_m$ denotes the integers modulo $m$. The neighborhood of a point $x \in \mathbb{Z}^d_{n_1 n_2 \cdots n_d}$ is given by the points

$$\mathscr{M}^d_r(x) = \{y \in \mathbb{Z}^d_{n_1 n_2 \cdots n_d} : |y_1 - x_1| \leq r, \ldots, |y_d - x_d| \leq r\}, \tag{11}$$

the Moore neighborhood of range $r$. The number of neighbors is

$$M_{r,d} = (2r+1)^d - 1. \tag{12}$$

In our model every agent ranks all of their neighbors in random order, and again we want to know $P_n$.

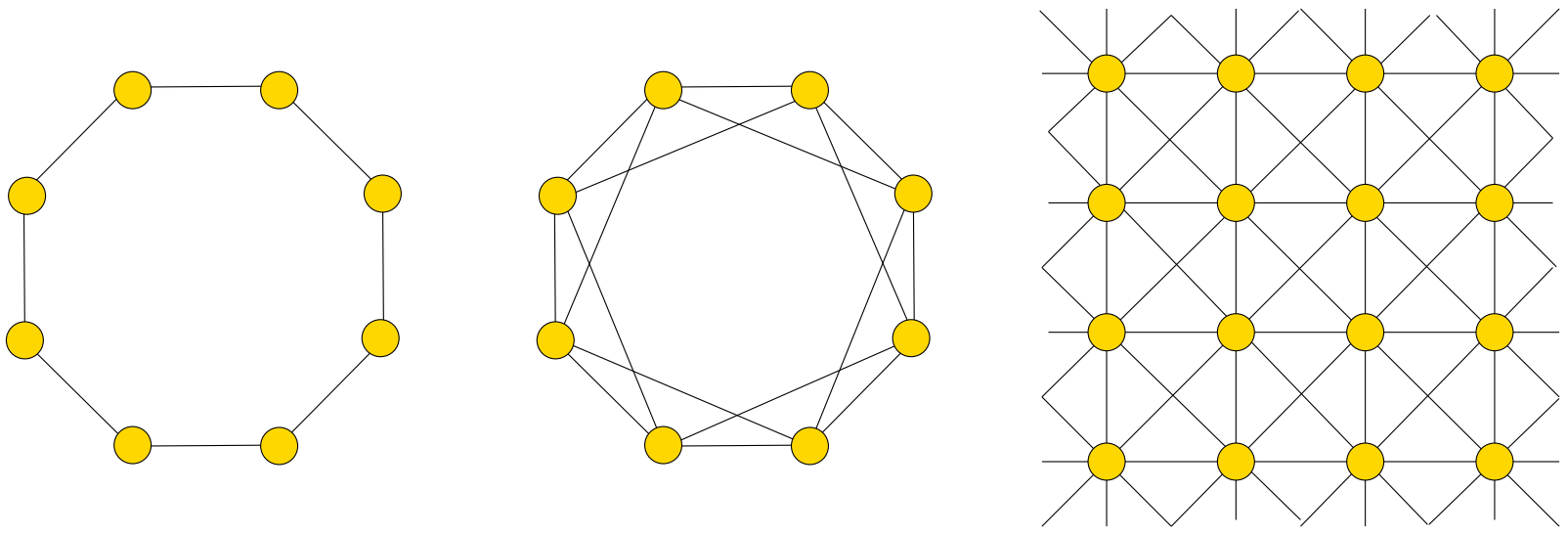

**Figure 5.** Grids: one dimensional with $r = 1$ neighborhood (a) and $r = 2$ neighborhood (b), two dimensional with $r = 1$ neighborhood.

The simplest lattice is the one dimensional grid with $r = 1$ neighborhood, also known as cycle graph $C_n$ (figure 5). This graph is bipartite for even $n$. For odd $n$, the only odd cycle that may appear in a stable partition is the cycle that includes all vertices. In order to prevent a stable matching the stability criterion (2) requires the alignment of *all* preferences to form an exclusive cycle (figure 2). Hence

$$P_n = \begin{cases} 1 & \text{if } n \text{ is even,} \\ 1 - 2^{-n+1} & \text{if } n \text{ is odd.} \end{cases} \tag{13}$$

For $r > 1$ and/or $d > 1$ the situation is different. We have many short odd cycles that may prevent a stable matching. The probability that three adjacent sites form an exclusive triangle is

$$\frac{2}{[M_{d,r}(M_{d,r} - 1)]^3} \tag{14}$$

for example. This probability is independent of $n$, but the number of triangles grows linearly with $n$. The same is true for exclusive and less exclusive cycles of larger length. If the probability that a random agent belongs to a stable, odd cycle is independent of $n$, we expect $P_n \simeq (1-p)^n$, and in fact this exponential decay is confirmed by simulations (figures 6 and 7).

**Conjecture 3** *The probabilty $P_n$ that a random instance of the stable matching problem on a grid admits a solution is*

$$P_n = \Theta(q^n) \tag{15}$$

*where $q < 1$ depends on the dimension $d$ of the lattice and the range $r$ of the neighborhood.*

From the numerical simulations we find that $q$ is close to 1 and depends only weakly on $r$ and $d$: $q = 0.97$ for $d = 1$ and $r = 2, \ldots, 5$ and $q = 0.98 \ldots 0.99$ for $d = 2$ and $r = 1, 2, 3$.

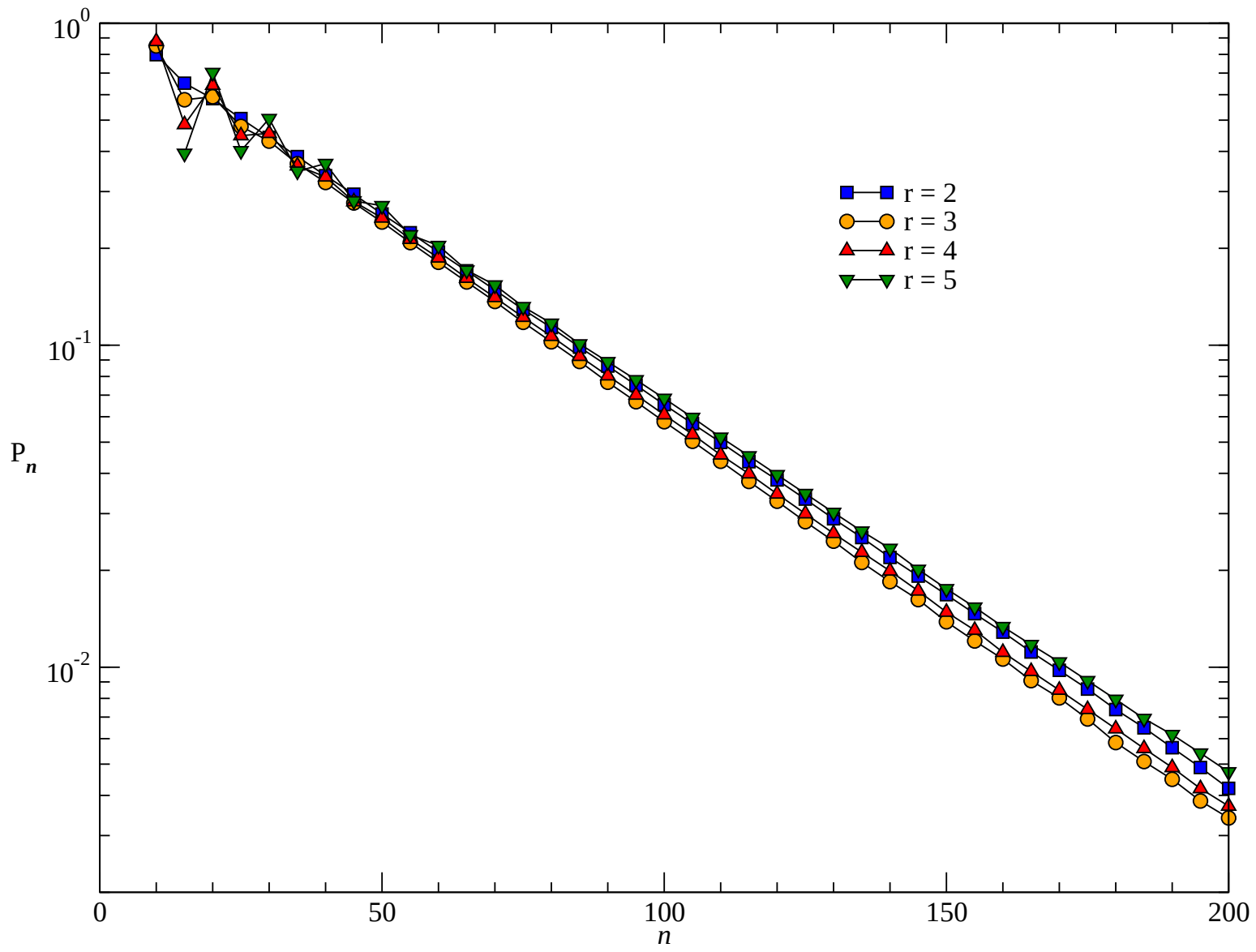

**Figure 6.** Probability $P_n$ of admitting a stable matching on 1$d$ lattices with $n$ vertices, periodic boundary conditions and varying range $r$ of the Moore neighborhood. Each data point represents an average over $10^6$ random samples.

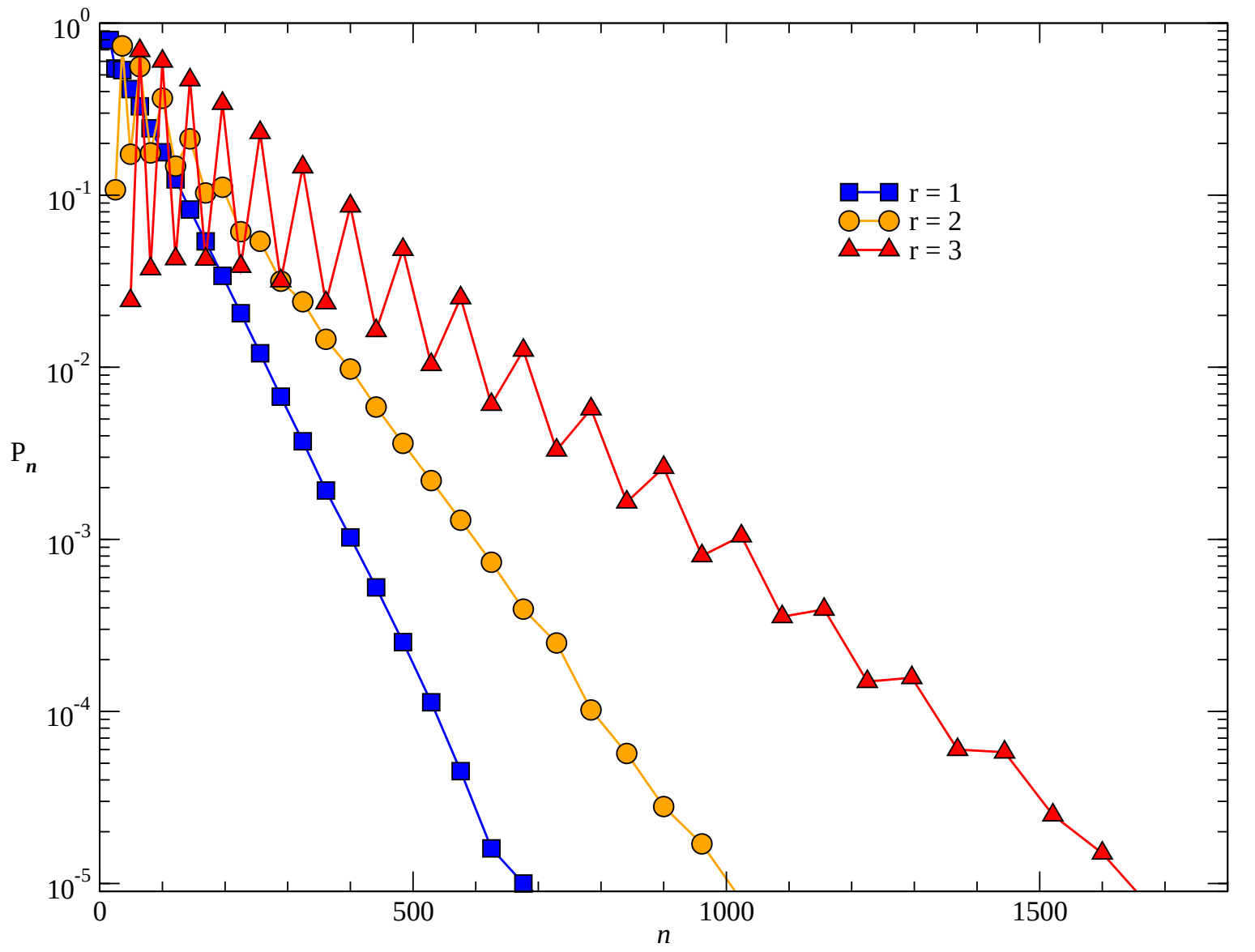

**Figure 7.** Probability $P_n$ of admitting a stable matching on 2$d$ lattices with $n = m \times m$ vertices, periodic boundary conditions and varying range $r$ of the Moore neighborhood. Each data point represents an average over $10^6$ random samples.

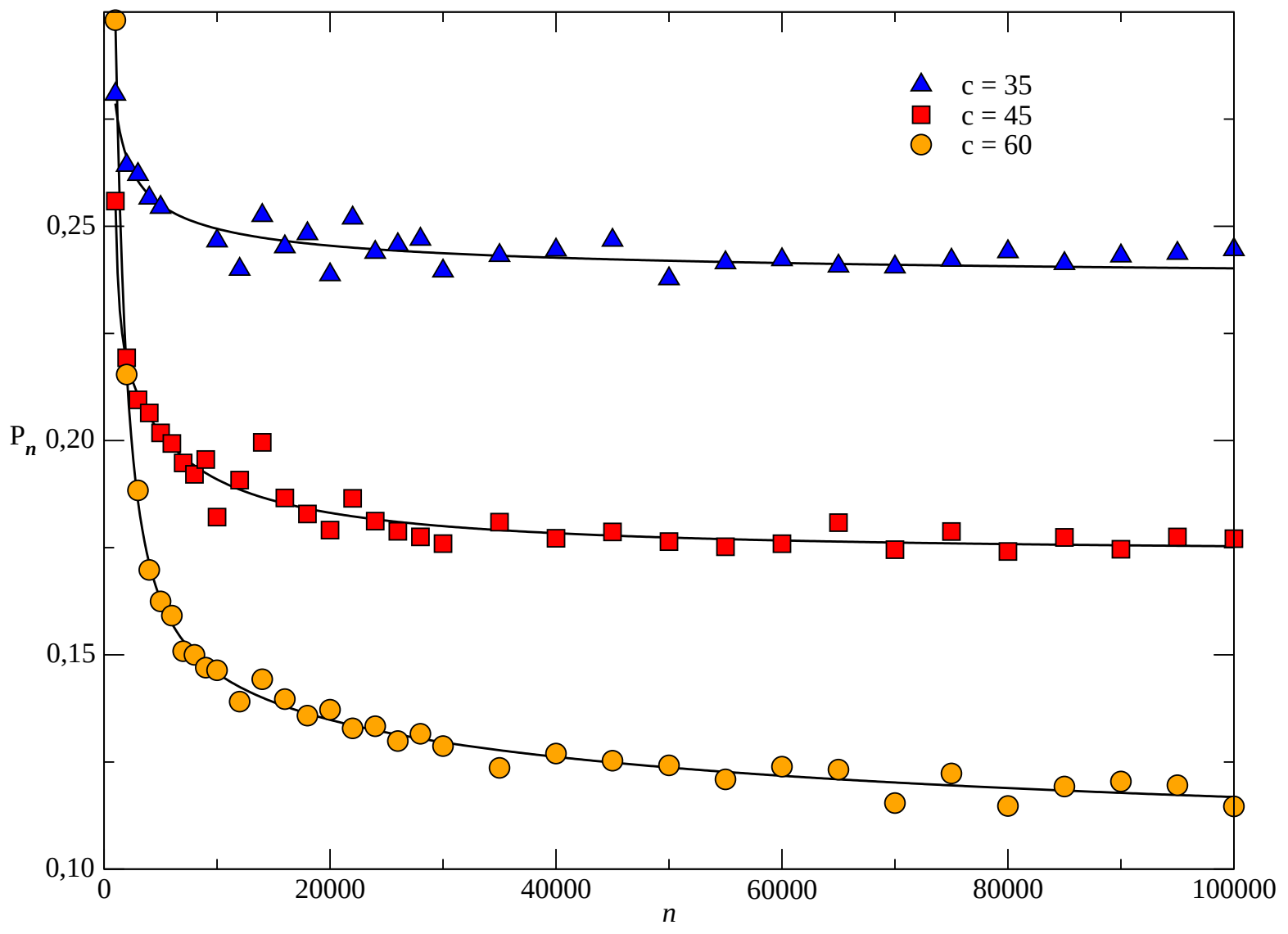

**Figure 8.** Probability $P_n$ of admitting a stable matching in finite Erdös-Renyi random graphs with average connectivity $c$. Each symbol represents an average over at least $10^4$ samples. The lines are two-parameter fits $P_n = a_1 + a_2/\sqrt{n}$.

### *3.3. Random Graphs*

In the last section we will discuss stable matchings in the the random graph ensemble $\mathbb{G}(n,p)$ introduced by Erdös and Rényi [17]. $\mathbb{G}(n,p)$ is the set of all graphs with $n$ vertices and an edge probability $p$, i.e. each pair of vertices is connected independently with probability $p$. If we scale $p = c/n$, the average connectivity of each vertex is $c$, i.e. we have a finite connectevity like in grids, but here the number of short cycles does not grow with $n$. The expected total number of triangles is $\frac{c^3}{3!}$ i.e. it is independent of $n$. The same is true for 5-cycles, 7-cycles etc.. Assuming again that the existence of a stable matching is determined by the absence of short odd cycles, we expect $P_n$ to be asymptotically independent of $n$ and bounded away form zero. The simulations (figure 8) suggest the asymptotics

$$P_n = P_\infty(c) + \mathcal{O}\left(n^{-1/2}\right), \qquad (16)$$

where the constant $P_\infty$ is an exponentially decreasing function of $c$ (figure 9).

**Conjecture 4** *Let $P_n$ denote the probability that a random instance of the stable matching problem on a random graph from $\mathbb{G}(n,c/n)$ admits a solution. Then $P_n$ converges to a number $P_\infty$ that depends on c,*

$$\lim_{n\to\infty} P_n = P_\infty(c) \qquad (17)$$

*and there exist numbers $\omega_1, \omega_2 > 0$ such that*

$$P_\infty(c) = \omega_1 e^{-\omega_2 c}. \qquad (18)$$

The simulations give $\omega_1 \approx 1.03$ and $\omega_2 \approx 0.042$.

For constant $p$ the average connectivity of a vertex is $pn$. Numerically we found that in this case $P_n$ behaves asymptotically like in the fully connected graph $p = 1$. This leads us to our last conjecture:

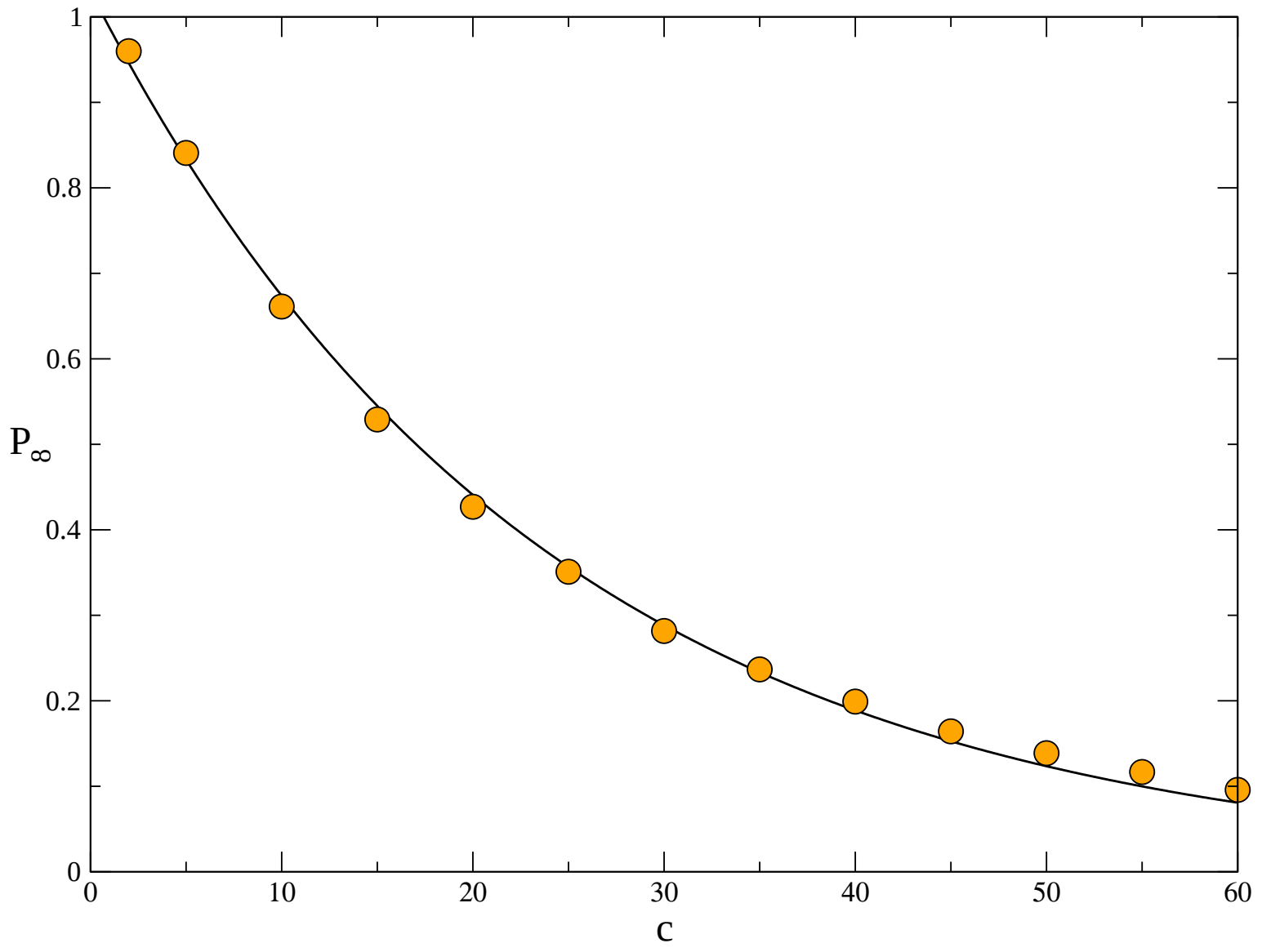

**Figure 9.** Probability $P_\infty$ of admitting a stable matching in the infinite Erdös-Renyi random graph with average connectivity $c$. Points are extrapolations from simulations in finite size graphs (see figure 8), the line is a two-parameter fit $P_\infty = \omega_1 \mathrm{e}^{-\omega_2 c}$.

**Conjecture 5** *The probability $P_n$ that a random instance of the stable roommates problem on a graph from $\mathbb{G}(n,p)$ admits a solution is asymptotically independent of $p$ and equals $P_n$ of the complete graph,*

$$P_n \simeq \mathrm{e}\sqrt{\frac{2}{\pi}}\,n^{-1/4}. \qquad (19)$$

## 4. Conclusions

We have studied the probability $P_n$ that a random instance of the stable matching problem admits a solution. Based on numerical simulations we conjectured the behavior of $P_n$ for the complete graph, for regular grids and for Erdös-Rényi random graphs. The conjectures are detailled enough to be falsified or verified by rigorous mathematical arguments in the future. The existence of a stable matching depends on the rareness of short cycles of low degree vertices. In grids there are plenty of them, leading to an exponential decay of $P_n$. In random graphs from $\mathbb{G}(n,c/n)$ the constant number of short cycles leads to a constant $P_\infty$ bounded away from zero. In finite connectivity graphs short cycles can easily coordinate to satisfy equation (2), but in $\mathbb{G}(n,p)$ graphs and the complete graph, this coordination is suppressed. At the same time we have a large number of short cycles, and the net effect is an algebraic decay of $P_n$.

## Acknowledgments

I am indebted to Boris Pittel for introducing me to this problem. I gratefully acknowledge useful discussions with Matteo Marsili and David Sherrington. This work was sponsored by the European Community's FP6 Information Society Technologies programme under contract IST-001935, EVERGROW, and by the German Science Council DFG under grant ME2044/1-1.